\documentclass[a4paper]{article}
\usepackage{graphicx}
\usepackage{chicago}
\newcommand{\be}{\begin{equation}}
\newcommand{\ee}{\end{equation}}
\newcommand{\bd}{\begin{displaymath}}
\newcommand{\ed}{\end{displaymath}}
\newcommand{\bea}{\begin{eqnarray}}
\newcommand{\eea}{\end{eqnarray}}
\newcommand{\bi}{\begin{description}}
\newcommand{\ei}{\end{description}}
\newcommand{\bq}{\begin{quote}}
\newcommand{\eq}{\end{quote}}
\def\i{\item}
\def\fo{\footnote}

\def\r{\rho}
\def\a{\alpha}

\def\G{\Gamma}

\def\t{\tau}

\def\om{\omega}
\def\Om{\Omega}

\def\s{\sigma}

\def\Z{{\sf Z\kern-4.5pt Z}} 
\def\R{{\sf R\kern-5.0pt I}}
\def\Q{{\sf C\kern-5.0pt Q}}

\renewcommand{\thefootnote}{\fnsymbol{footnote}}
\begin{document}
\bibliographystyle{chicago}
\author{Alexander~Unzicker\\
 Pestalozzi-Gymnasium  M\"unchen\\
{\small e-mail: aunzicker@lrz.uni-muenchen.de}\footnotemark[1]}
\title{Galaxies as Rotating Buckets - a Hypothesis on the \\
Gravitational Constant Based on Mach's Principle}
\date{September 7, 2003}
\twocolumn[
\maketitle

\begin{abstract}
According to Mach's principle inertia has its reason in the presence
of all masses in the universe. Despite there is a lot of sympathy for 
this plausible idea, only a few quantitative frameworks have been proposed
to test it. 
In this paper a tentative theory is given which is based on Mach's 
critisism on Newton's rotating bucket. Taking this criticism seriously,
one is led to the hypothesis that the rotation of our 
galaxy is the reason for gravitation.
Concretely, a functional dependence of the gravitational constant 
 on the size, mass and angular momentum of the milky way is proposed
that leads to a {\em spatial\/}, but not to a temporal variation of $G$.
Since Newton's inverse-square law is modified, flat rotation curves of galaxies can be
explained that usually need the postulate of dark matter. 
While the consequences for stellar evolution are discussed briefly,
a couple of further observational coincidences are noted 
and possible experimental tests are proposed. 

\end{abstract}
\vspace{1.0cm}]

\section{Introduction}

I start\footnotetext[1]{Postal address: 
Melssheimerstr.11, D-81247 M\"unchen, Germany}
 summarizing briefly some experimental and observational facts in 
gravitational physics which are not fully understood yet: dark matter,
dark energy, the pioneer anomalous acceleration and the discrepant
$G$ measurements. In section~2, Mach's principle and in particular
the rotating bucket filled with water, is discussed in detail.
The reader will find there a step-by-step motivation using heuristic arguments
and thought experiments that led to the central hypothesis which is
summarized at the end of the section. 
In section~3, the current observational evidence and all further
possible tests of the theory are outlined. A discussion
of open theoretical problems and an outlook is found in section~4.

\renewcommand{\thefootnote}{\arabic{footnote}}

\subsection{Dark matter} \label{Dar}
Doppler-shift measurements allow to determine the velocity of gas clouds
orbiting around 
spiral galaxies. These velocities $v$ were expected to show a Keplerian 
behaviour $v^2 = G M/r$, $r$ being the distance to the galactic center,
i.e. do decay with $r^{-1/2}$ outside the range 
of visible matter. Contrarily to that, a large number of galaxies
show flat rotation curves (\shortciteNP{MFB}; \citeNP{Per}), i.e. constant velocity $v$ 
far out the visible radius. This region being called halo, the standard
explanation postulates a large amount of non visible (dark) matter 
inthere. However, there is not only disagreement whether dark matter is
predominantly cold (bayronic) 
or hot, but no satisfactory observational evidence for the postulated 
objects or particles in laboratory physics.
Even if such an evidence will occur some day, further surprising
observational facts, for example the Tully-Fisher relation or the
recently discovered M(BH)-$\s$ relation \cite{Fer}
have still to be explained. For an overview on dark matter in spiral galaxies,
see \citeN{Bos}. 

As an alternative, various modifications of the Newtonian distance
law have been proposed. While many of these models can be ruled out by
the rather sophisticated data availible (for a good review on 
observational constraints see \shortciteNP{Agu:01a}), a modification of Newtonian
dynamics (MOND, for a review see \citeNP{Mil}) has been the most sucessful until now.
While a `fundamental
acceleration' in the order of $a_0=\frac{c^2}{R}$, ($R$ being the radius 
of the universe) appears in MOND, there is no theoretical framework consistent
with the `rest of physics' backing that theory \cite{Sco}.

In summary, dark matter should be still regarded as a phenomenon not
fully understood rather then a cornerstone of a standard 
cosmological model.

\subsection{Dark energy}

While the interpretation of Hubble's distance-redshift
relation from 1929 was still disputed in the 1950s, the discovery of 
the cosmic microwave background (CMB) left no doubt
that the  universe is expanding. While theorists were taking for granted 
that the expansion caused kinetic energy to be transformed in potential one,
 it was uncertain for a long time which part prevailed, 
i.e. if the average density was high enough to stop 
the expansion and to start a recontraction (closed universe, $\Om>1$), or if 
the universe would expand forever because of $\Om<1$. Since 
the total observed density is at least {\em close\/} to $1$ regarding
orders of magnitude, as a consequence of the evolution of the universe 
in promordial times $\Om$ must have been as close as $10^{-60}$ to $1$.
For that reason, many theorists believe that $\Om$ is {\em exact\/} $1$,
i.e. the universe is completely flat. 

The numerical value of the Hubble constant $H$ suffered from %
a huge uncertainty for a long time. While most of the methods
yielded a value around $80 \, km s^{-1}/Mpc$, the analysis of 
high-redshift 
supernovae Ia suggested a value near $50 \, km s^{-1}/Mpc$.
Since the increasing presicion of the different methods left no doubt,
the most reasonable conclusion from the datasets was that the Hubble
constant increased with time (The actual precise value is
$71.4 \, km s^{-1}/Mpc$, \citeNP{Ben}), i.e. the expansion of the universe is 
{\em accelerated\/} (\shortciteNP{sup}; \shortciteNP{sup2}; \citeNP{sup3}; 
\citeyearNP{sup5}; \citeyearNP{sup4}).


To explain this data, the standard cosmological model assumes 
a density with {\em repulsive\/}
instead of attractive properties called dark energy (DE). According to
the latest WAMP measurements \cite{Ben} 
its relative amount is $0.73$, yielding with dark and visible matter ($0.23$ and $0.04$)
altogether $\Om \approx 1$.

While this argument is numerically correct, one must raise the question
if the above concept of defining $\Om$, according to which 
kinetic energy transforms into Newtonian gravitational energy, is tenable. 
Independently of 
 $\Om$ being smaller or greater than $1$, one should expect a {\em decrease\/}
of $H$ with time, which is excluded by observation. Given the fact that
dark and visible matter are attractive and dark energy is repulsive, encompassing both
fractions in $\Om$ seems to be an ill-defined addition of different 
quantities (in agreement, \citeNP{Sta}). 

Instead of being happy with two forms of matter (DM, DE) to which
no counterpart in the laboratories has been detected yet, 
cosmological data should raise doubts on a simple law on 
energy conservation that uses the Newtonian potential energy.


\subsection{The Pioneer anomalous acceleration}

The two spacecrafts Pioneer 10 and 11 that left the solar 
system\fo{The data availible is in the range of 20-50 AU.}, show
an unmodeled, approximately constant acceleration $a_p$ of about 
$ 8.7 \cdot 10^{-10} m \, s^{-2}$ towards the sun \shortcite{And}. Despite much
effort, no systematic reason has been found yet. There is also 
a diurnal and annual residual acceleration not yet explained.
While the authors still favour an undetected systematic error as an
explanation for
the anomalous acceleration, they admit that this would require some 
unlikely coincidences. As in the case of MOND, the numerical coincidence
of $a_p$ with $c^2/R_U$ has launched a couple of speculations (\shortciteNP{And}, 
p.~43~ff.). To avoid the enormous difficulties of filtering out $a_p$
from noisy data, further space missions are currently proposed \cite{Nie}. 


\subsection{Discrepant $G$ measurements} \label{Dis}

A couple of years ago, discrepant measurements of Newton's constant
$G$, which led to a raise in the 
CODATA uncertainty attracted attention \cite{Uza}. 
Still recently, the most precise values of \citeN{Gun:00} and \shortciteN{Qui},
which both used sophisticated versions of the torsion balance method,
are discrepant. Some measurements are obtained by using superconducting
gravimeters \shortcite{Bra}. Interestingly, the accuracy of these instruments
has been pushed to about $10^{-10} m \, s^{-2}$ - the same order of
magnitude as $a_0$.

In general, some attention is given to the short-range 
of measurements of Newton's law \cite{Ade}.
On the other hand, few knowledge on $G$ can be obtained at the large scale.
Even if a {\em time\/} variation of $G$ seems to be ruled out
(see the excellent review on the change of physical constants,
 \citeNP{Uza}, p.~25ff.),
there is little observational evidence that constrains a {\em pure spatial\/}
variation.
While the atomic spectra of astrophysical data would immediately
detect even slight variations of the `atomic' constants like $c$, $\a$ and $\hbar$,
it quite difficult to detect slight spatial variations of $G$ on a galactic scale.


I have listed - with a decreasing experimental evidence from 
section~\ref{Dar} to \ref{Dis} - some observational problems 
which are quite well-known.
Gravitational physics has however some unsatisfactory aspects
from the theoretical point of view I shall address in the next section.

\section{A new aspect of Mach's principle: galaxies as rotating buckets~?}

\subsection{Arbitrary elements in Newton's theory}

While Einstein claimed to be inspired by the idea of Mach, 
it has not been included into general relativity 
(\citeNP{Einst:17}; \citeNP{Bon}). 
From the Schwarzschild metric of a spherical mass distribution it is
obvious that GR does not `care about' distant masses, since it would
take an identical form if the rest of the universe was empty.

Though not explicitly stated by him,
Mach's principle is is nowadays known as follows:
The reason for inertia is that a mass is accelerated {\em
with respect to all other masses in the universe\/}.
The strenght of gravitation can therefore be related to the mass distribution 
of the universe. A recent proposal using this paradigm is Barbour's 
\citeyear{Bar:02} scale invariant theory, an overview on other
Machian attempts can be found in \citeN{BaP}.

Mach critizied in particular Newton's concept of absolute 
space. In his famous example of the rotating bucket filled with 
water\fo{for a very didactical presentation, see \citeN{Wil}.},
Newton deduced the existence of an absolute, nonrotating space
from the observation of the curved surface the water forms.
Mach criticized this `non observable' concept of absolute space as follows,
suggesting that the water was rotating {\em with respect to masses at large
distance\/}:
`No one is competent to say how the experiment would turn out if the sides of the
vessel increased in thickness and mass till they were ultimately
 several... [miles]... thick.' 

Particulary clear is the formulation of the same problem given
by \citeN{Sci:53}: `Using Mach's principle we can predict that 
the angular velocity of the Earth, as deduced from a local
dynamical experiment (such as the motion of a Foucault pendulum),
will be the same as that deduced kinematically from the apparent motion
of the fixed stars.'

 Since in Newton's theory there is no causal connection
between these facts, Sciama considers this choice of inertial frames
as one of two arbitrary elements in Newton's theory, the other one
being the value of the gravitational constant $G$.


In summary, one must conclude that either an important theoretical concept
is still missing or the determination of inertial frames by the
fixed stars and Foucault pendulum is just a meaningless coincidence.

In the next paragraph, I will outline how a particular proposal
for a quantification of Mach's principle can be related to the
observation of dark matter. Then, I will discuss a new
aspect of the rotating bucket.


\subsection{A first approach to Mach's Principle} 

When following the ideas of Mach, one raises the question how the 
masses of the universe could influence the gravitational interaction. 
The most general possibility is a functional dependence of Newtons constant $G$:

\be
G= G(m_i, \vec r_i) \label{fu}
\ee
Since the unit $\frac{m^3}{s^2 kg}$ seems 
to contain $c^2$, 
this leads us to consider
terms of (\ref{fu}) with the unit $\frac{m}{kg}$, as the constant
$\t = \frac{c^2}{8 \pi G}$  used in general relativity. 
Since $\sum_{i}  \frac{r_i}{m_i}$ is obviously not reasonable,
the next apparent guess is

\be
\frac{c^2}{G} =\sum_{i}  \frac{m_i}{r_i}. \label{sum}
\ee
Of course, other forms of (\ref{fu}) could still reflect
Mach's principle. One of the most interesting theories in this
direction has been proposed by Barbour (\citeyearNP{Bar:02}, sec.~7).
The proposal (\ref{sum}) has first been developed by
\citeN{Sci:53}  in the context of a modified gravitational 
{\em potential\/}, 
\be
\varphi= G \, \sum_{i} \frac{m_i}{r_i},
\ee
from wich he deduced\fo{\citeNP{Sci:53}, eqn.~6, with a sign change.} 
$\varphi= -c^2$.
An application of this idea to the dynamics of topological defects
will be discussed by \citeN{Unz:03b}. There is however another promising
possibility, namely to insert 
(\ref{sum}) directly into Newton's law. The potential
instead can then be brought to the form
\be
\varphi = -c^2 \log \sum_{i} \frac{m_i}{r_i}   \label{pot}
\ee
from which follows Newton's law
\be
\label{New}
-\nabla \varphi = -c^2 \frac{\sum_{i} 
\frac{m_i \vec r_i}{r_i^3}}{\sum_{i} \frac{m_i}{r_i}} := -G(m_i, r_i) \sum_{i} \frac{m_i \vec r_i}{r_i^3}.
\label{for}
\ee
 Since the distant masses give the biggest
contribution to the sum, $G$ does practically not change on small
scales and creates the illusion of an $1/r^2$- law.
The form (\ref{pot}) of the gravitational potential is however particulary interesting,
since logarithmic potentials shows on the {\em large\/} scale a $1/r$-decay of the
gravitational field that is suitable for describing flat rotation 
curves of galaxies. Indeed, if we consider a spherical mass 
distribution with homgeneous density $\r$,
\be
\sum_{i} \frac{m_i}{r_i} = \int{V} \frac{\r dV}{|\vec x - \vec r(V) |} = \frac{M}{r} \label{Sum}
\ee
holds for every point $x$ outside the sphere, $r$ being its distance to
the center; this is well-known from potential theory. 
Thus outside the sphere (\ref{for}) reduces to 
$c^2 \frac{M}{r^2}/ \frac{M}{r} = c^2/r$, yielding
a $1/r$-decay of the the force.

The however big problem is that using (\ref{for}), the flat rotation curves would appear
outside a mass distribution of the size of the {\em whole universe\/}
 and {\em not\/} at the scale of a galaxy.
Furthermore, due to the expansion of the universe, (\ref{sum}) would
predict a relative increase of $G$ which is of the order of $dt/t_H$,
the Hubble time, which is definitely ruled out by observations 
\cite{Uza}.
Contrarily, the size of {\em galaxies\/} remains pretty constant at
least for the present epoch. This leads to the hypothesis to apply the
structure of (\ref{sum}) to a galactic scale.
Indeed, the expression 
\be
\frac{v^2}{G} =\sum_{galaxy}  \frac{m_i}{r_i} \label{sumgal}
\ee
could yield the correct order of magnitude if $v$ is a velocity 
characteristic for the galaxy. There are basically two options:
$v$ could be the Hubble velocity measured with the dipole anisotropy
of the CMB, but rotation curves do apparently not depend on that.

The other option is that $v$ is of the order of $v_{max}$, the
maximum rotational velocity of the galaxy. 
Even if (\ref{sumgal}) could in principle predict flat rotation curves,
it cannot be the correct expression yet,
since the varying distance earth-sun due to the elliptical orbit would
cause a huge annual signal in $G$.

In a certain sense the dependence on such a velocity
 seems even trivial, because $v_{max}$ is the 
usual measure of the mass of the galaxy by 
equating the gravitational and centripetal force. 
This equality may be however of deep nature since it is a rotational
velocity as $v_{max}$ that acquires a surprising property in the
rotating bucket thought experiment, once we try seriously to
derive quantitative consequences from it.
This will be investigated in the next section.

\subsection{The bucket-galaxy analogy}

In his famous criticism on Newtons rotating bucket filled with water,
Mach's implicit conjecture was that the water surface even in a rotating 
bucket would remain flat, given that the `walls' approach the size
of the universe or, if we refer to the preceeding paragraph,
to the size of a galaxy. If we suppose that the huge bucket can still be perceived
as rotating from a test mass at the outside (distant galaxy or quasar),
Mach's statement has a dramatic implication.

Given that outside the bucket it appears as a rotating coordinate
system, {\em what force\/} should act as centripetal force that keeps
the water surface flat~? If inertial and rotating frames are equivalent,
by what transformation can we obtain equations of motion~?
One posssible quantification of Mach's
principle is that the compensating centripetal force needed here {\em is
gravitation\/}. In this case the need of considering rotating frames 
as inertial ones  generates gravitation as a consequence of that transformation.
I will try to develop this hypothesis with a slightly
modified version of Newton's bucket. 

\subsection{Universe consisting of two point masses.} \label{Uni}

Let's analyze a gedanken (`thought') experiment with two masses $m_1$ and $m_2$ in
a beyond empty universe, and let's further assume that they rotate around their center
of mass (COM).\fo{Let's assume that time and length measurements are possible. There is 
an interesting idea of of \citeN{Bar} that relates the perception of time
to the change configurations. According to that idea, time would not even pass
in this thought experiment.}
According to Newtonian mechanics, the centripetal force $F_Z$ must be 
equal to the gravitational force $F_G$:
\be
 \frac{m_1 m_2}{m_1+m_2} \frac{v^2}{r} =G \frac{m_1 m_2}{r^2}, \label{red}
\ee
$r$ being the distance between the two masses\fo{The reduced mass appears
at the l.h.s.}. 

From a Machian point of view, it makes no sense to speak about a rotation,
because there is no absolute space and the two masses define the inertial 
coordinate frame. Thus there a two possible solutions:\fo{A similar situation 
has  been analyzed by \citeN{Chu}.}

a) since there is no measurable rotation,
the two masses start moving towards each other due to gravitational attraction.

b) since the situation is indistinguishable from the rotation, the outcome
should be the same: no change in the relative distance $r$. This means: when
there is {\em no rotation\/}, there is {\em no gravitation\/} either.

Even if the two masses cannot measure it, a rotation could
not be excluded in principle, a case in which a) offers no solution. No
stable orbits in the two-body problem would be possible.

For that, solution b) seems to be much more logical, since even in a Machian context 
the classical mechanics of rotating systems is valid and need the concept
of a centripetal force. b) suggests however that $F_Z$ and $F_G$ are identical.
The rotation in case b) would just take the role of a (non observable)
gauge field. 

\subsection{Non-measurable tangential velocities}

The deeper reason for this somewhat surprising interpretation is that it
is impossible to measure {\em tangential\/} velocities instantaneously. While
radial velocities are usually measured by the redshift of atomic spectra,
no similar objective method exists for tangential velocities. What {\em
can\/} be measured is the change in position relative to other masses,
but as long as the distance does not change, this is indistinguishable from 
a rotation of the coordinate system of the observer.
Thus in the above case there is no way to measure tangential velocities,
and no way to measure whatever rotation.

\subsection{The influence of distant masses}

\begin{figure}[h]
\includegraphics[width=70mm]{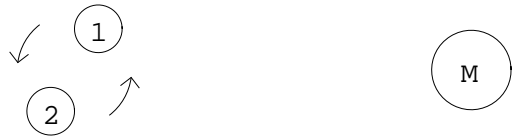}
\caption{The additional mass $M$ at a large distance influences the possible dynamics
of $m_1$ and $m_2$}
\label{dreimass}
\end{figure}
We continue now the gedanken experiment by placing a third mass $M$ at a 
constant distance $R$ from the COM of $m_1$ and $m_2$ (see sketch
fig.~\ref{dreimass}). To maintain the analogy with the universe
($M$ as distant galaxy or quasar), we shall assume that
${R \gg r}$ and ${M \gg m}$, but in a way that both ${M/R \gg m/r}$ and 
${M/R^2 \ll m/r^2}$ holds\fo{This implicates also negligible tidal forces.},
 even if only the very first condition is
essential. The appearance on $M$ has a couple
of consequences how $m_1$ and $m_2$ `perceive the universe':
\bi
\i 
It makes sense now to speak about a rotation of $m_1$ and $m_2$,
when their sight lines crosses the `quasar' $M$.
\i
While $m_1$ and $m_2$ alone had a `gauge freedom' chosing any
rotating coordinate system up to a angular velocity correponding
to $\om_{max}= c/r$, the constraint is now $c/R$, which limits
their possibility to consider themselves as {\em nonrotating\/}.
That means, if $v$ exeeds $c \frac{r}{R}$, they cannot `gauge away'
their rotation any more  by assuming $M$ rotating.
\i
When rotating, $m_1$ and $m_2$ may determine their mass ratio by 
the excentricity of their COM, i.e. when the minimum (maximum) 
distance from  $m_1$ to $M$ is different from that one of $m_2$.
Without $M$, this was impossible.
\i
Taking the option b) of the previous section, $m_1$ and $m_2$
{\em need a centripetal force\/} if they want to keep $r$ constant,
a necessity that is induced by $M$.
\i
The central hypothesis of this proposal is that gravitation is
generated by this necessity and perceived as a $1/r^2$ attraction
law that depends on the product of the attracting masses.
\i
Gravitation is however {\em not\/} an illusion (when talking about
two masses we still could see it like that), because three masses 
prohibit the description in a {\em global\/} inertial system.
\i
Since it is not illusory, gravitation acts with radial symmetry
also in the third dimension, while it was `primarily invented' 
to act towards a rotation axis. I will come back to this point 
in section~\ref{DE}
\i
If we equate the `induced' centripetal force with the Newtonian\fo{If 
$m_1$ and $m_2$ conclude that their interaction is
Newtonian and if they suppose this Newtonian law is universal,
they may deduce that $M$ and the COM of $m_1$ and $m_2$ have to
rotate around the COM of all masses in order to keep $R$
fixed. This is however, a generalization that cannot be measured.}
force, the same condition as (\ref{red}) holds, but the gravitational
force cannot be `gauged away' any more:
\be
G (m_1+m_2) = v^2 r \label{sum2},
\ee
which is apparently similar to eqn.~(\ref{sumgal}).
\ei

\subsection{Generalizing the three-masses-thought experiment}

\label{3mass}

We are seeking now a general formula that contains (\ref{sum2})
as a special case and yields the functional dependence of $G$.
There are three problems to be addressed:
how to generalize, $m$, $r$, and $v$.

\begin{figure}[h]
\includegraphics[width=70mm]{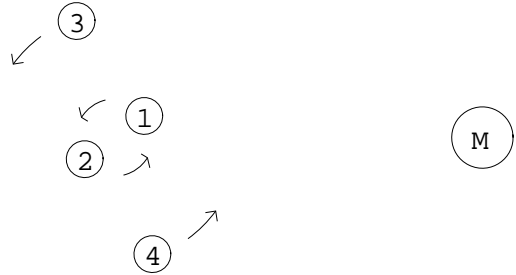}
\caption{If we place additional masses $m_3$, $m_4$ which rotate 
around their COM, all masses $m_1 - m_4$ contribute to the gravitational 
interaction.} 
\label{fivemass}
\end{figure}

\paragraph{Masses.}
We observe that with Newtonian interaction, the motion of $m_1$ and $m_2$ 
is not quite affected by the presence of $m_3$ and $m_4$, while $m_3$ and $m_4$ feel 
well the presence of $m_1$ and $m_2$. Therefore, in a rotating system, 
the sum  should be taken over the masses inside a given radius only. 
This limiting radius must enter the seeked formula, thus one cannot uphold
eqn.~(\ref{sumgal}).

\paragraph{Velocity.}
Given that the inner and the outer pair of masses need not to orbit with
the same velocity, a reasonable `medium rotating velocity' has to
be determined. One must also consider that there should be a cancelation
of clockwise and counterclockwise rotating masses; otherwise $G$ would
immediately depend of the temperature of a star. This excludes the 
possibility of introducing a term like $\sum_i v_i^2$.
The natural measure for the rotation of a system of mass points is
the angular momentum $\vec L$. Thus, a reasonable way to define
a medium {\em tangential\/} velocity $\bar v$ is
\be
\bar v^2 = \frac{\vec L^2}{I M} = 
\frac{(\sum_i m_i \vec r_i \times \vec v_i)^2}{\sum_i m_i r_i^2 \sum_i m_i},
\label{neuv}
\ee
when $I$ is the moment of inertia, 
$\vec r_i$ is the distance vector from the rotation axis,
and the sum is taken over all the masses of the rotating system 
for which $r_i < r$ holds.
Thus, $r$ being the distance from the rotation axis, 
\be
G(r, r_i, m_i, v_i)= 
 \frac{r \vec L^2}{I M^2} \label{neuG}
\ee
is the proposal I favour and shall discuss in the next section. 

I should address here however some unsatisfactory
theoretical aspects of (\ref{neuG}).
Technically, one may replace $\sum_i m_i r_i^2 \sum_i m_i$ in the
denominator of (\ref{neuv}) by $(\sum_i m_i r_i)^2$, which yields
the correct units as well.  

More displeasantly, 
the `sum over all masses of the rotating system' is not really 
a well-defined quantity. It is the angular momentum as ingredient
 of the sum that should tell if a system is rotating, not an
a-priori selection of masses. The reason for this poor logic
is that in the above thought experiment the distant mass $M$ 
(`Quasar') plays a decisive role without entering eqn.~(\ref{red}), 
respectively~(\ref{sum2}). Paradoxically, the physics of the
three-masses situation is neither changed by the amount of $M$ nor
by its precise distance $R$, as long as $R \gg r$ holds. How 
should $M$ and $R$ enter any formula if there values are not important~?
$M$ just `communicates' the frame of inertia without playing any further 
role. It seems however that 
one inevitably runs into this paradox once we take Mach's
criticism of Newton's bucket seriously.\fo{However, even 
summation or integration over space coordinates we are used to, 
could turn out to be ill-defined, since space
without matter is meaningless in a Machian sense.}
To remedy this, eqn.~(\ref{neuG}) took $M$ out `by hand', 
and should be for that regarded only as a first tentative to arrive
at quantitative predictions. 

\subsection{Rotating subsystems} \label{sub}

While fig.~\ref{fivemass} is a primitive model of the universe if 
we consider $m_1$ - $m_4$  as our galaxy and $M$ as the rest, our galaxy
contains subsystems like the solar system that rotate with respect 
to the (already rotating) galaxy at a distance $r$ from the rotation
axis. 
Following the arguments given above, it is impossible for the solar 
system to consider itself as nonrotating, not even with respect to  
the milky way. Consequently, this additional rotation with respect 
to the galaxy should create an extra force. While its structure must be
again Newtonian, one may express this new effect by a correcting term 
that analogous to (\ref{neuG}) but containing the angular momentum, the size 
and the mass  of the  solar system. Similar arguments hold for 
subsequent systems like  planet-moon etc.

The definite quantitative form of the correction has to be still derived.
The inner systems are compelled to consider themselves rotating by 
the outer ones. If we return to the argument in section~\ref{Uni},
there is however a `gauge freedom' left that allows us to assume
the boundary of the universe to move with a {\em tangential\/}
velocity $c$. One may also suggest that there is a connection of
the corresponding centripetal acceleration $c^2/R_U$ that seems to
be detected at the boundaries of the respective subsystems
($a_0$ at galaxies, $a_p$ at the solar system).  

\subsection{Conclusion}

I summarize here the assumptions and collect the formulas for
quantitative predictions.

The milky way and galaxies in general are regarded 
as `Newton's buckets' in which the Machian
hypothesis holds. Since there must be a centripetal force, 
gravitation is generated by this necessity. 
Thus one may construct a functional dependence of $G$ in a way that
$F_Z$ and $F_G$ have the same value for a orbital motion in the 
milky way. $G$ on a given distance $R$ from the axis of rotation
depends therefore on the angular momentum and mass distribution
of the galaxy:
\be
G(r, r_i, m_i, v_i)= \frac{r \vec L^2}{I M^2}=\frac{\bar v^2 r}{M}=
\frac{r (\sum_i m_i \vec r_i \times \vec v_i)^2}{\sum_i m_i r_i^2 (\sum_i m_i)^2},
\label{neuG2}
\ee
while the sums are taken over the masses $m_i$ with velocities $v_i$ and
distances from the rotation axis $r_i$, whereby  $r_i < r$ must hold. $r$ is again
the distance from the rotation axis of the position where $G$ is measured.
$\bar v$ represents a characteristic velocity of the galaxy.

\section{Observational and experimental tests}

\subsection{Reproduction of Newton's law}

Eqn.~(\ref{neuG2}) contains $\vec L$, $M$ and $r$ as variables. If we assume
in a very crude approximation the milky way as a rigidly rotating flat disk
with constant density, $M \sim \bar v^2$ holds, thus $G$ increases lineary
with the distance from the center. If $r_0$ is the distance of the solar
system from the center of the milky way, the variation is however too
small to be perceived at the scales of celestical mechanics,
and even smaller for $1/r^2$-gravity tests on earth.

\subsection{Rotation curves and dark matter}

Since for gas clouds in the halo of a galaxy very large distances $r$
occur in Newton's law $a=G \frac{M}{r^2}$, where $M$ is the total mass of the galaxy,
the functional dependence (\ref{neuG2}) causes the acceleration $a$
following the law
\be
a=\frac{\bar v^2 r}{M} \frac{M}{r^2} = \frac{\bar v^2}{r}, \label{acc}
\ee
where $\bar v$ is fixed by (\ref{neuv}) and should have the same
order of magnitude as $v_{max}$. 
The $1/r$ decay of the acceleration
obviously predicts a rotation curve which is flat outside the (visible)
mass distribution. 
So far, no nonvisible mass distributions have to be postulated. The
evidence for a relatively higher amount of dark matter in low-luminosity-galaxies
is adressed below (\ref{ste}). Since this proposal does not arbitrarily
modify Newton's potential just in the galactic plane, it seems not
to be in conflict with the further observational constraints listed by 
\shortciteN{Agu:01a}.

\subsection{Variation of $G$ and its consequences for stellar evolution}
\label{ste}

When talking about a variation of $G$, almost all proposals in the
literature deal with a {\em time\/} evolution of $G$. Against this popular idea
originated by \citeN{Dir:38a} an overwhelming observational evidence has been 
collected \cite{Uza}. 
In agreement with this facts, this proposal predicts no time evolution,
since the quantities $\vec L$, $M$ and $R$ of galaxies are constants
at least in the present period.

However, it predicts both a huge {\em spatial\/} variation of $G$ 
inside a galaxy and different values of $G$ for each galaxy, e.g. for
each rotating mass distribution.
This raises the question how the apparent uniformity of star
populations matches
this prediction. In particular, one has to investigate how the 
mass-luminosity-relation and correlated quantities like stellar lifetime
is affected by a different $G$\fo{There could be even an 
influence on cepheid-based distances measurements.}.

\citeN{Tel}, in criticizing Dirac's idea, derived a $G^7$-dependence
of the luminosity $L$. While this is approximately correct for 
a star with given mass under temporal variations of $G$, the argument
does not apply to the question what kind of stars would be formed under
conditions with a spatially different $G$\fo{A similar error would be to 
to confuse the issues of the mass-luminosity-relation for a 
single star with the mass-to-light
ratio of an ensemble of stars.}.

For main sequence stars I consider here for simplicity, $R \sim M$ 
\cite{Shu}\fo{On the upper main sequence, $R \sim M^{0.6}$.} 
is the condition for stars to exist. This reflects the 
fact that the inner temperature $T_i$ is determined by parameters
of the nuclear reactions 
and $T_i \sim G \frac{M}{r}$ due to the virial theorem\fo{To
be precise the preconditions 
for the virial theorem have to be reanalyzed for the present
proposal, see \citeNP{LanI}.}.

Thus $G \frac{M}{r} = const.$ should approximately hold for any main sequence.
Consequently, stars that we observe in the inner parts of
the milky way, where according to (\ref{neuG2}) $G$ is lower,
should have  either a smaller radius or a greater mass than
in standard models. Since size rather than mass should determine
the instability region in the HR-diagram, one would expect 
the latter possibility and a mass-to light-ratio $\G$ proportional to
$G^{-1}$. Even if there is still a huge uncertainty
regarding $\G$, this prediction seems to
hit the right direction.

As a consequence, large high-luminosity galaxies
with a lower medium value of $G$ would show a higher
mass-to-light ratio. This is in agreement with the observational
fact that low-luminosity galaxies seem to be relatively more 
dominated by dark matter.

\subsection{The Tully-Fisher-relation} \label{TFR}

Both  conventional dark-matter models and alternative 
post-newtonian potentials do not explain the luminosity
dependence $L \sim v^4$ discovered by Tully and Fisher.

Since the dependence $G \sim R$ should increase $G$ with $v$,
(\ref{neuG2}) influences again the mass-to-light ratio.
Therefore, an increase of $L$ with a higher power of $v$ 
is predicted as in conventional models. (\ref{neuG2})
is therefore approximately in agreement with observation.

\subsection{Globular clusters}
 
Due to the difficulties outlined in section~\ref{3mass}, unfortunately
it is not clear how to apply eqn.~(\ref{neuG2}) to globular clusters 
which are located in in the halo. Given that a spatial variation of $G$
is expected for the disc, it is at least possible that $G$ may have 
different values at the cluster positions. Since clusters are 
nonrotating objects, there should be a discrepancy towards lower $G$
if any. Though a detailed theoretical study is needed, a considerable
change of the HR-diagram would be the consequence, as indeed it is
observed. Since a lower $G$ raises the Chandrasekhar limit 
\be
M_{C} \approx (\hbar c)^{\frac{3}{2}} G^{-\frac{3}{2}} m_p^2,
\ee
($m_p$ being the proton mass and $\hbar$ Plack's constant),
much less supernovae can occur, which is in agreement
with the low metallicity 
of population II stars. The apparent stability
and absence of gravitational contraction is currently explained
with gravothermal oscillations, which seems to be a complicated and
non obvious mechanism. 

\subsection{The Pioneer anomaly}

I outlined in section~\ref{sub} the possible influence of rotating
subsystems in a galaxy, like the solar system. Qualitatively,
eqn.~(\ref{neuG2}), calculated with the parameters of the solar
system, predicts a slight increase of $G$ with distance from the center
of rotation (the sun) which results as an extra acceleration towards
the sun. The large contributions of Jupiter and Saturn to the 
angular momentum of the solar system suggest that an extra acceleration
due to an increased $G$ could be `switched on' at that distance,
as indeed  the form of the diagram in \citeN{And}, p.~19, shows.
A preliminary quantitative analysis however, predicts an effect which
is too big by several orders of magnitude. This could indicate that
(\ref{neuG2}) has to be replaced by a similar formula.


\subsection{$G$ and absolute $g$ measurements}

The next level of a rotating subsystem to which a possible correction
has to be applied is the earth-moon-system; in a more general sense,
even free fall experiments on earth are subject to these possible
corrections. Given the considerable discrepancies in the $G$ measurments
it would be desirable if parameters like exact time and position on
earth 
would be recorded.

Given that the proposed model predicts a {\em spatial\/} variation
of $G$ that depends on the distance from the galactic center,
an annual signal of $G$ due to the earth orbit is expected.
Since the relative amplitude is $1 AU/R_g \approx 6 \cdot 10^{-10}$,
an amplitude of 6 nGal in $g$ with maximum in Summer should be observed.
Absolute $g$-measurements of the earth's gravity field can be done
by superconducting gravimeters (SG) that reach a precision of about 
$0.1 \, nm \, s^{-2}$.
Unfortunately the earth has lots of noisy signals in its field,
starting from the difficulties in modeling nutation effects up to
atmospheric and hydrological disturbances. Currently, there is an
unmodeled signal/uncertainty of about 20 nGal \cite{Kro}. However, the
worldwide linking of SG data in combination with the excellent data
of the beginning GRACE mission should make such a signal 
detectable in the future.

\subsection{The accelerated universe and dark energy}
\label{DE}
In the vicinity of rotating galaxy, a $1/r$-decay of the gravitational
field is predicted, which means a stronger field up to a scale 
of galaxy halos. However, since the largest rotating 
structures are galaxies, the angular momenta of different 
galaxies will cancel out on larger scales. Therefore, this proposal
predicts {\em no gravitational interaction at all\/} for 
intergalactic, cosmological scales. Distant galaxies and quasars
are only `used' as frame of reference (as the distant mass $M$ in the
thought experiment) that however influences the appearance of
gravity in the described Machian picture.

Given that a couple of observations may have to be reinterpreted
following this Machian idea, there is at least no obvious
contradiction with the observed high-redshift supernovae. 
Contrarily, a constant non decelerated Hubble expansion could 
be understood in this picture without postulating a new form
of matter called dark energy.

Rather than postulating a repulsive gravity that has not shown
up elsewhere in physics, one may interprete 
the high redshift supernova data in terms of a 
{\em violation of energy conservation\/} on cosmological scales.
It is well-known that energy conservation can be deduced from a 
radially symmetric force field. Inverting that theorem, one deduces
that {\em if energy conservation is violated, the force field
cannot obey radial symmetry any more\/}. This absence of radial
symmetry on intergalactic scales is actually what appears in a Machian 
interpretation of the rotating bucket and what is included 
in eqn.~(\ref{neuG2}).

Regarding the the observed self-similar fractal structure of galaxy
distributions one should keep in mind that phase transitions in 
thermodynamics that show similar structures are satisfiable 
modeled with a next-neighbour interaction and therefore do not require 
a long-range interaction; in this aspect, the present proposal is in
agreement with observation.

\subsection{Galaxy evolution and radial symmetry}

At the heart of this proposal stands the relation between
gravitational and centripetal forces. While gravitation obeys
radial symmetry, rotational forces are directed towards an axis
and act, so to speak, in two dimensions. If rotation indeed generates
three-dimensional gravity, a homogeneous distribution of matter in the 
early universe, for which we have evidence from CMB, should be contracted
along one dimension, because centrifugal forces may compensate only two.

Observing the shape of disc galaxies one must raise the question why most the
universe is dominated by structures that violate radial symmetry so
obviously, 
if Newton's radial law with constant $G$ is really the only interaction
that matters.
Despite much progress in detail, we do not really understand why galaxies
have the shape and size they have. Is it a coincidence that the surface
of all galaxies matches approximately\fo{Assuming milky way as one of $10^{11}$
standard galaxies with $r=10 \,kpc$ and $R=1.3 \cdot 10^{26} m$ matches up to a 
factor~2.} the surface of a sphere with radius of the universe~?

It is well-known that galaxies, clusters and superclusters show
a hierarchical, selfsimilar structure \cite{Man}. Recently, the fractal
dimension of that structure has been measured to $2 \pm 0.2$ \cite{Ros},
promoting  the conjecture that the universe has a fractal dimension of two. Why
should a radially symmetric interaction produce such a particular form~? 
Once we have evidence that energy conservation is violated on large scales,
there is no need to keep radial symmetry as a dogma.

\section{Theoretical Problems}

I collect briefly some unconventional implications of this proposal
that have to be investigated with more stringent methods than the
heuristic arguments given in section~2.

\paragraph{The modification of the gravitational constant $G$}
 leads to a 
force that depends on the position {\em and\/} velocities
of other masses. Moreover,
the spatial dependence does refer to the distance from the 
center of rotation, not to the distance to a given point $\vec R$.

\paragraph{Energy conservation.}
The spatially dependent $G$ in (\ref{neuG2}) obviously 
disagrees with Newton's 3rd law on large scales. As a consequence,
energy conservation in its conventional form is violated on large
scales (see section~\ref{DE}). In view of the high-redshift 
supernovae, this theoretical challenge that has to be taken 
however anyway. 

Since the Newtonian potential energy seems to do not a good job on
intergalactic scales, a different Lagrangian has to be derived
that allows to deduce equations of motion from a variational
principle. Eqn.~(\ref{neuG2}) suggests that velocities and
vector valued quantities may enter that Lagrangian.

\paragraph{Relation to the universe.}
Since this proposal discarded all versions of Mach's principle that
included {\em all\/} masses in the universe\fo{The rest of the universe
had only the meaning of a distant frame of reference like the mass
$M$ in fig.~\ref{dreimass}.} and focussed on the effects
on a galactic level, one may raise the question what meaning
remains for the `fundamental acceleration' $c^2/R_U$ that actually
appears. This has to be understood investigating galaxy-formation
processes, as well as the coincidence 
$\frac{c^2 R_U}{M_U} \approx \frac{\bar v^2 R_G}{M_G}$.

\paragraph{The compatibility with general relativity} 
remains entirely to be clarified. 
Only one point is satisfying yet:
The Machian interpretation of the Newton's bucket seems to
be in agreement with the equivalence principle, since  masses
do react on inertia and gravitation in the same way, and 
the two interactions are in a sense equated  by definition.

\paragraph{Outlook.}
It is hard to be happy with the paradox situation that distant
masses create the necessity of gravity without their parameters
$M$ and $R$ showing up in the formalism. Given this
logical difficulties, It may be justified to discard Mach's 
comment on the rotating bucket. If one concludes to do that,
he should believe however in absolute space.

I consider it as a nice feature of the proposal outlined here that it does not 
introduce further arbitrary parameters into Newton's theory but tries
to get rid of the arbitrary parameter $G$. Deviating from the law of energy
conservation - even at large scales - seems however a too expensive toll. 
Thus much more theoretical understanding has to be achieved
before one can claim the validity of such an approach. 
Given the enormous riddles astronomers are observing with increasing precision,
this proposal may however be worth to be tested in detail.

\paragraph{Acknowledgement.}

Hannes Hoff prevented me from giving up this idea at an early stage, 
and contributed important arguments. This work wouldn't have  been
possible either without the uncountable inspiring discussions with Karl Fabian.
Comments of Ettore Minguzzi are acknowledged.

\end{document}